\documentclass[reprint]{revtex4-1}
\usepackage{amsmath}
\usepackage{graphicx}
\usepackage{mathtools}
\newcommand*\binconj[1]{\bar{#1}}
\newcommand{\kron}{\otimes}
\newcommand{\ket}[1]{| #1\rangle} 

\begin{document}

\title{Superdense Coding Interleaved with Forward Error Correction}
\author{Ronald J. Sadlier}
\affiliation{Quantum Computing Institute, Oak Ridge National Laboratory, Oak Ridge, TN, USA}
\author{Travis S. Humble}
\affiliation{Quantum Computing Institute, Oak Ridge National Laboratory, Oak Ridge, TN, USA}

\begin{abstract}
Superdense coding promises increased classical capacity and communication security but this advantage may be undermined by noise in the quantum channel. We present a numerical study of how forward error correction (FEC) applied to the encoded classical message can be used to mitigate against quantum channel noise. By studying the bit error rate under different FEC codes, we identify the unique role that burst errors play in superdense coding, and we show how these can be mitigated against by interleaving the FEC codewords prior to transmission. We conclude that classical FEC with interleaving is a useful method to improve the performance in near-term demonstrations of superdense coding.
\end{abstract}

\maketitle

\section{Introduction}
Entanglement is a versatile resource that enables many new methods of  communication including teleportation, quantum secret sharing, and superdense coding \cite{Imre2005, Wilde2013}. Superdense coding (SDC), in particular, enables a sender Alice to transmit a two-bit message to a receiver Bob by sending only one member of a shared bipartite entangled quantum state \cite{Bennett1992}. This is twice the classical capacity expected from direct transmission of a single unentangled qubit. In addition to increased capacity, correlations within the entangled state provide the added benefit that an eavesdropper cannot recover the message by simply intercepting the single transmitted particle. These unique features of SDC have been demonstrated by several different experiments and shown to provide both greater channel capacity and security relative to direct communication techniques \cite{Mattle1996,Li2002,Barreiro2008,Chiuri2013,Liu2015}.
\par
A limitation on SDC performance in realistic settings is the presence of channel noise, which corrupts the transmitted state. Channel noise reduces the rate that information is reliably transmitted and has the potential to completely undermine the expected benefits from SDC \cite{Bennett1999}. Previously, Shadman et al.~have investigated the classical capacity for SDC when transmitting in the presence of depolarizing noise  \cite{Shadman2010,Shadman2011,Shadman2013}. Their results quantified the decrease in capacity for single-sided and double-sided transmission, in which noise acts on one and both particles, respectively. Notably, Shadman et al.~found a crossover point in the depolarizing noise model for which double-sided SDC does no better than direct quantum transmission. This cross-over point represents the conditions for which the classical SDC capacity was smaller than when using direct quantum transmission. 
\par
The presence of noise in experimental realizations of SDC raise the question as to how this type of communication can benefit from the use of error correction techniques. Both quantum and classical error correction codes could in principle offer protection to the transmission of information. Quantum error-correction (QEC) codes use ancillary qubits to construct a higher-dimensional code space that can detect errors in the logical code words \cite{Gaitan2008}. There have been several experimental demonstrations that show the benefits of QEC for protecting against sources of channel noise, including demonstration using optically encoded qubits \cite{Pittman2005,Bell2014}. However, the requirement of using additional ancilla is at odds with experimental limitations on the number of simultaneously generated photon states, which make it difficult to prepare codeword states for existing QEC codes.  
\par
Classical forward error correction (FEC) offers an alternative to QEC that is more easily realized by existing quantum communication technology. In this setting, FEC codes rely on classical ancilla to redundantly encode the state of an input bit string. When used in conjunction with SDC, the resulting classical codewords are encoded into quantum states and transmitted across a noisy quantum channel. These received states are detected and then decoded into classical strings representing the original message. An immediate advantage of this approach is that it does not require additional entanglement in quantum resources. The use of FEC codes also does not require changes to existing quantum hardware for implementation, but rather can integrate with the software systems that manage existing hardware \cite{HumbleSadlier2014}. 
\par
However, the noisy quantum channels that lead to errors in the received codewords are not always equivalent to well-studied classical channels, and the performance of an FEC code for SDC depends on the details of the quantum channel. Indeed, this point has been emphasized in recent work by Boyd et al.~ who found that binary codes are limited in the capacity that is achievable \cite{Boyd2015}. This limitation was traced back to the distinction between quantum states as soft symbols and the measured bits as hard symbols. Boyd et al~have shown that non-binary classical error correction codes can offer better performance and may be viewed though the lens of a $d$-ary memory-less classical model for entanglement-assisted communication \cite{Boyd2015}. 
\par
In this paper, we investigate the performance of several binary FEC codes for mitigating errors from superdense coding over noisy quantum channels. In particular, we calculate the bit-error rate (BER) for finite-length messages over a Pauli channel using numerical simulations of the transmission. We calculate the decoded BER for the repetition, Hamming, and Golay FEC codes in the presence of depolarizing noise. We then investigate the use of data interleaving to mitigate against the structure of these Pauli operator errors. We present results from numerical simulations of these different system designs and we identify the relative performance under the Pauli noise model.
\par
The remainder of the paper is organized as follows. In Sec.~\ref{sec:sdcreview}, we review the superdense coding protocol and present a model for noisy transmission. In Sec.~\ref{sec:fec}, we introduce notation for FEC codes and provide an example of how FEC coding affects the bit-error rate. In Sec.~\ref{sec:simulations}, we present numerical simulation results for FEC encoded SDC, and in Sec.~\ref{sec:interleaving} we analyze the impact of interleaving for protecting entangled pair states from some Pauli errors. Our conclusions are then presented in Sec.~\ref{sec:conclusions}.
\section{Noisy Superdense Coding}
\label{sec:sdcreview}
The superdense coding (SDC) protocol begins by assuming that users Alice and Bob share a pair of qubits prepared in a known entangled state. We will consider this state to belong to one of the four Bell states, 
\begin{equation}
\begin{split}
& \ket{\Phi_{AB}^{(\pm)}}= \frac{1}{\sqrt{2}}(\ket{0_A,0_B}\pm \ket{1_A,1_B}) \\
& \ket{\Psi_{AB}^{(\pm)}}= \frac{1}{\sqrt{2}}(\ket{0_A,1_B}\pm \ket{1_A,0_B}) \\
\end{split}
\end{equation}
where $0$ and $1$ denote the eigenstates of the Pauli $Z$ operator and $A$ and $B$ label the different subsystems \cite{Nielsen2011}. For concreteness, let the shared initial state be $\Phi^{(+)}$ and the corresponding initial density matrix be defined as $\rho^{(+)}$. Alice uses this fiducial state to transmit a two-bit message $b$ to Bob by encoding the binary coefficients $b_0$ and $b_1$ of the message into the state. The message is encoded by applying the unitary operator
\begin{equation}
\label{eq:oa}
\mathcal{O}_{A}(b) =\mathcal{O}_{A}(b_0,b_1) = X_{A}^{b_0} Z_{A}^{b_1}
\end{equation}
to subsystem $A$. Each of the four possible operators maps the initial state uniquely into one of the maximally entangled Bell states and we will use the shorthand notation $\rho_{b} = \mathcal{O}_{A}(b)\rho_{AB}^{(+)}\mathcal{O}_{A}(b)^{\dagger}$ to denote the state encoding the message $b$.
\par
After encoding the message, Alice transmits subsystem $A$ to Bob, who performs a projective measurement in the Bell basis to identify the state prepared by Alice \cite{BarencoEkert1995}. The complete set of Bell-state measurements may be modeled by the set of projection  operators $\Pi(b) = \rho_b$, where $b  = \{0,1,2,3\}$ uniquely labels the measurement outcome. The probability of measuring Bell state $b'$ given the encoded value $b$ is given as
\begin{equation}
\mathrm{Prob}(b'|b) =\textrm{Tr}[\Pi_{b'} \rho_{b}].
\label{eq:measurement}
\end{equation}
For the noiseless protocol, $\mathrm{Prob}(b'|b) = \delta_{b',b}$, and
Bob may translate the detected state to recover the two-bit message according to map generated by Eq.~(\ref{eq:oa}). Thus, in the absence of noise, each two-bit message is faithfully transmitted using a pure state and the capacity of each channel use is two bits.
The idealized SDC protocol neglects several sources of noise that may be encountered in a realistic implementation. First, the preparation of the initial state shared by Alice and Bob may deviate from the protocol specification. In particular, it may be represented by a nonmaximally-entangled mixture of bipartite states. This deviation from the protocol may be due to environmental induced decoherence of the initially pure state or to noisy preparation of the entangled pair. Second, Alice's encoding operations may be faulty from imprecise application of the Pauli operators $X, Z,$ and $XZ$. This also leads to less than maximally entangled Bell states with the effect that the resulting measurement statistics will not be perfectly correlated with the input message. Third, the transmission of subsystem $A$ from Alice to Bob may encounter channel noise that introduces errors in the amplitude and phase of the state. Additional possible errors include loss, whereby the transmitted particle is not received by Bob, and leakage, in which the particle is perturbed into a state not detected by Bob's measurement apparatus.
\par
We will consider a model of noisy SDC given by the one-sided Pauli channel
\begin{equation}
\mathcal{E}(\rho_b) = \sum_{j=0}^{3}{p_j \mathcal{O}_{A}(j) \rho_b \mathcal{O}_{A}(j)^{\dagger}} 
\label{eq:noise}
\end{equation}
with $p_0 \in [0,1]$ and $p_{1,2,3} \in [0,1/3]$ constrained by $\sum_{j}{p_j} =1$ and the operator $\mathcal{O}_{A}(j)$ defined in Eq.~(\ref{eq:oa}). The quantum channel $\mathcal{E}$ serves as an effective model to account for a variety of noise sources including transmission noise, state preparation noise, and detection errors. Equation (\ref{eq:noise}) also has the property that it maps every Bell-state to a mixture of Bell states and,  consequently, this is a covariant channel with respect to the Pauli basis defined by Eq.~(\ref{eq:oa}). We will limit subsequent analysis to the case $p_0 = 1-p$ and $p_{1} = p_2 = p_3 = p/3 $ for $ p \in [0, 1]$, which is known as depolarizing noise limit for noise parameter $p$ \cite{Nielsen2011}.
\par
While the probability to receive the incorrect quantum state is $1 - p_0$, the probability to receive the incorrect classical message is slightly less. This is because the message encoded in a received Bell state will have either  one, two, or no bits corrupted. Given the output state of the channel in Eq.~(\ref{eq:noise}), the probability that Bob observes the message $b'$ when Alice sent the message $b$ is formally
\begin{equation}
\mathrm{Prob}_\mathrm{SDC}(b'|b) = \mathrm{Tr}[\Pi_{b'} \mathcal{E}(\rho_b)]
\label{eq:noisymeasurement}
\end{equation}
For the depolarizing channel, the corresponding transition probabilities are given as
\begin{equation}
\mathrm{Prob}_\mathrm{SDC}(b'|b) =
\begin{cases} 
1-p & b'_0b'_1=b_0b_1 \\
\frac{p}{3} & b'_0b'_1=\binconj{b_0}b_1 \\ 
\frac{p}{3} & b'_0b'_1=b_0\binconj{b_1} \\
\frac{p}{3} & b'_0b'_1=\binconj{b_0}\binconj{b_1}
\end{cases}
\label{eq:prob_sdc}
\end{equation}
with $\binconj{b_j}$ the binary conjugate of $b_j$.  The bit error rte (BER) measures the ratio of incorrect bits to the number of bits transmitted. For the one-sided SDC depolarizing channel, the frequency with which the binary message is incorrectly decoded is given by
\begin{equation}
\label{eq:bersdc}
\mathrm{BER_{SDC}} = \frac{2p}{3}
\end{equation}
\par
As a point of comparison, consider the case that Alice and Bob use direct quantum transmission instead of entanglement for communication. Alice then uses the noisy quantum channel twice to transmit a single qubit correctly with probability $p$. However, the action of the Pauli $Z$ operator changes only the phase of the qubit and, therefore, does not lead to an incorrect bit. The corresponding transition probabilities for Bob to successfully recover both of Alice's bits are
\begin{equation}
\mathrm{Prob}_\mathrm{Direct}(b'_0b'_1|b_0b_1) =
\begin{cases} 
(1-\frac{2p}{3})^2 & b'_0b'_1=b_0b_1 \\
\frac{2 p}{3}(1-\frac{2p}{3}) & b'_0b'_1=\binconj{b_0}b_1 \\ 
(1-\frac{2p}{3})\frac{2p}{3} & b'_0b'_1=b_0\binconj{b_1} \\
\left( \frac{2p}{3}\right)^2 & b'_0b'_1=\binconj{b_0}\binconj{b_1}
\end{cases}
\label{eq:prob_direct}
\end{equation}
In this case, the BER also evaluates to 
\begin{equation}
\mathrm{BER_{Direct}} = \frac{2p}{3}
\end{equation}•
\par
Although the BER for SDC and direct transmission are the same in the case of a depolarizing channel, the classical channel capacities are not. In particular, the classical capacity for SDC in the presence of depolarizing noise has been given previously as \cite{Bennett1999}
\begin{equation}
C_\mathrm{SDC}(p)=2+(1-p)\log{(1-p)}+p \log{\frac{p}{3}}
\end{equation}
while the corresponding capacity for a single use of direct transmission is
\begin{equation}
C_\mathrm{Direct}(p)=1+(1-p)\log{(1-p)}+p \log{p},
\end{equation}
which is less than half of the former. 
\par
Finally, it is useful to compare the capacity for superdense coding with that of a 4-ary classical communication system. Both communication methods are capable of transmitting 4 distinct symbols, although SDC leverages the non-local correlations between Alice and Bob for this purpose. Bennett et al.~have previously shown that the capacity for both systems is the same under the symmetric channel \cite{Bennett1999}.
\section{SDC with Forward Error Correction}
\label{sec:fec}
Forward error correction (FEC) codes increase the reliability of transmission over a noisy channel by encoding the message such that a transmission contains redundant information. The receiver may detect a transmission error and correct it without the involvement of the sender, making forward error correction codes useful on simplex (i.e., one-way) communication channels. For a SDC channel, these codes are applied to the classical message before being encoded to quantum symbols. In addition, the measured classical symbols by Bob are then decoded according to the FEC code. An FEC code with distance $d$ can detect up to $(d-1)$ errors in the received message and correct any message with $\left \lfloor{(d-1)/2}\right \rfloor$ errors or less.
\par
Formally, we will represent an $[n,k,d]$ code by the $n$-by-$k$ generator matrix $\mathbf{G}$, which transforms an input binary vector $\mathbf{b}$ into a codeword $\mathbf{c}$, i.e.,
\begin{equation}
\label{eq:gen}
\mathbf{c} = \mathbf{G} \mathbf{b}
\end{equation}
encodes $k$ input bits into an $n$-bit codeword $\mathbf{c}$. These $n$ bits will then be transmitted pairwise using the SDC protocol. If $n$ is odd, we concatenate two consecutive codewords for transmission of a $2n$-bit message. Alternatively, the codeword may also be padded with an extra bit. In either case, consecutive bit pairs are mapped into the Bell states according to Eq.~(\ref{eq:oa}). Applying FEC to the original classical message increases the size of the transmitted information by a factor $n/k$, the inverse of the code rate. An input message $\mathbf{b}$ of size $2k$ is encoded by FEC as the codeword vector $\mathbf{c}$ of size $2n$ and then transmitted by the ensemble of states
\begin{equation}
\rho_{\mathbf{c}} = \prod_{i=0}^{n-1}O_{A}(c_{2i},c_{2i+1})\rho_0 O_{A}(c_{2i},c_{2i+1})^{\dagger}
\end{equation}
where $c_i$ is the $i$-th bit as defined by Eq.~(\ref{eq:gen}).
\par
Detection of $n$ quantum states sent by Alice leaves Bob with the received message $\mathbf{d} = \mathbf{c} + \mathbf{e}$, where $\mathbf{e}$ represents the error imparted by the noisy channel. Forward error correction codes will permit decoding of the error provided its weight (Hamming distance) is within the bound previously mentioned. If so, then the error can be decoded using the parity matrix $\mathbf{H}$ for the $[n,k,d]$ code, which satisfies 
\begin{equation}
\mathbf{b}' = \mathbf{H} \mathbf{d}.
\end{equation}
Whenever $\mathbf{H} \mathbf{d} \neq 0$, an error has occurred. However, the error can be identified only if the weight is less than the distance of the code.
Thus, the probability for a given bit pair to be correctly decoded is given by the number of ways in which a correctable error may be received by Bob. 
\par
For example, an FEC code with distance $d=3$ corrects up to a single error in the received codeword. From Eq.~(\ref{eq:prob_sdc}), the corresponding probability  that the original message is correctly received following error correction under the depolarizing channel is the bit error rate (BER)
\begin{equation}
\mathrm{BER}_\mathrm{SDC} = (1-p)^n + \frac{2p}{3} (1-p)^{(n-1)},
\label{eq:prob_sdcfec}
\end{equation}
which measures the number of incorrectly decoded codewords. The BER is useful for assessing the impact of the FEC as it measures errors remaining after error correction. It also permits errors to be isolated to the individual codewords that contain them.
\begin{table}
\caption{Properties of FEC codes tested}
\begin{tabular}{ | c | c | c | }
\hline Coding & Distance & Rate \\
\hline No FEC & 0 & 1 \\
\hline Repetition $[n=\{3,5,7\}, k=1]$ & $\frac{n}{2}$ & $\frac{k}{n}$ \\
\hline Hamming $[n=7, k=4]$ & 1 & $\frac{4}{7}$ \\
\hline Golay $[n=24, k=12]$ & 3 & $\frac{1}{2}$ \\
\hline
\end{tabular}
\label{tab:codes}
\end{table}
We have tested the performance of SDC when using several different FEC codes to protect against depolarizing noise. We will present numerical results of these studies in the next section but we first provide an example using the Hamming [7, 4, 3] code. Consider the explicit example that Alice sends Bob the 4-bit message $\mathbf{b} = \{1,0,0,1\}$ using the Hamming $[7, 4]$ code. The FEC encoder using the Hamming [7, 4] encoding prepares the codeword $\mathbf{c} = \{0,0,1,1,0,0,1,0\}$, where we have appended an extra 0 to make the codeword even. These 8 bits are then sent to the quantum encoder, which translates the each pair into a Bell-state symbol, i.e., $\Phi^{(+)}, \Psi^{(-)}, \Phi^{(+)},$ and $\Psi^{(+)}$. 
\par
During transmission, each of these states will experience noise according to the probabilistic model given by Eq.~\ref{eq:noise}. The effect of the noise depends on both the state and the error operator. For example, an $X$ error will transform $\Psi^{(+)}$ to $\Phi^{(+)}$ and Bob's measurement results will be mapped to the bit pair $(0,0)$ according to Eq.~\ref{eq:oa}. This binary error can be corrected however using the Hamming code. Assuming no other errors during the transmission of four consecutive Bell states, the complete message that Bob receives in this example is $\{0,0,1,1,0,0,0,0\}$. Following decoding of the received message, the original message is recovered as $\{1,0,0,1\}$ and the resulting BER is 0. The FEC succeeds in this example because the single binary error resulting from corruption of the transmitted quantum codeword is correctable when using the Hamming code. Similarly, any other weight-one binary error in the message can be corrected following measurement of the complete encoded message. However, weight-two and higher binary errors cannot be corrected by the Hamming code. Moreover, these errors may be mistakenly characterized as weight-one errors and the resulting attempt to correct the message will corrupt it further. It is notable that the binary errors in the received message trace back to the quantum noise operations that act on the transmitted Bell state. 
\section{Numerical Simulations of Noisy SDC Transmission}
\label{sec:simulations}
Numerical simulations of the noisy SDC transmission model presented in Sec.~\ref{sec:sdcreview} were used to investigate the influence of FEC codes on BER. We use the QITAKT framework for software-defined quantum communication systems that has been presented previously \cite{HumbleSadlier2014}. Briefly, the QITKAT library is an extension of the GNU Radio real-time signal processing framework that adds primitives for quantum information applications. The GNU Radio framework provides a run-time manager and a large variety of conventional communication methods that can be easily integrated into a graphical work flow and deployed to interact with hardware components \cite{url:gnu-radio}. We leverage the GNU Radio framework using the QITKAT library to construct quantum communication applications that can be either deployed directly to hardware or simulated using numerical methods \cite{Pooser2012}. In the present study, we use numerical simulation methods to investigate the behavior FEC codes under the Pauli noise model.
\begin{figure}
\centering
\includegraphics[width=1.0\columnwidth]{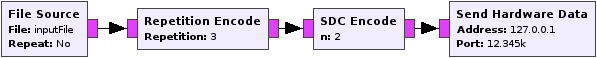}\\
\vspace{2em}
\includegraphics[width=1.0\columnwidth]{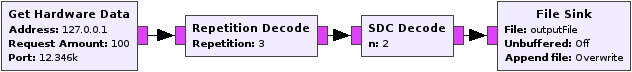}
\caption{(top) Alice's \textsc{qitkat} application implements FEC and SDC encoding steps before pushing data to the middleware layer to transmit the specified Bell states. (bottom) Similarly, Bob's application gets measurement results from the receiver middleware layer and performs the SDC and FEC decoding operations. }
\label{fig:sdcflowgraph}
\end{figure}
\par
Example QITKAT applications are shown for Alice and Bob in Fig.~\ref{fig:sdcflowgraph}. These applications represent pre-processing steps that prepare messages to be sent to the communication middleware layer. For Alice's application, an input file source feeds a stream of bits to the FEC code block (the repetition code is shown in the example). The output from the FEC encoder passes through the SDC encoder, which converts each pair of consecutive bits to a transmission symbol that is then sent to the quantum transmitter. The quantum transmitter consists of both hardware that prepares and measures the entangled quantum states as well as middleware that manages these processes. In our case studies, we have replaced hardware instances with a numerical simulator that interprets the middleware control signals and prepares a numerical representation of the quantum state \cite{HumbleSadlier2014}. The simulator uses a model of the quantum channel to generate a representation of the measurement results that is observed by Bob's receiver. The receiver middleware interprets the measurement results and performs decoding of the  transmission. As shown in Fig.~\ref{fig:sdcflowgraph}, the decoding steps mirror the encoding performed by Alice.
\par
We use a numerical simulator that takes advantage of the algebraic structure of both the Bell states and the Pauli channel for noisy transmission. In particular, we model Bob's detectors as projective measurements in the Bell basis as given by Eq.~(\ref{eq:noisymeasurement}). These measurements are therefore equivalent to sampling of the Pauli channel. We then use a Monte Carlo method to sample the simulated transmission and chose a specific outcome for each measurement. The sampling is biased according to the error rates $\{p_x, p_y, p_z\}$ with random instances drawn using a pseudo-random number generator. For the symmetric depolarizing noise, we consider the noise parameter $p$ within the range $[0, 0.1]$ as suggested by recent experimental results \cite{PhysRevLett.96.190501}. 
\par
\begin{figure}[ht]
\centering
\includegraphics[width=1.0\columnwidth]{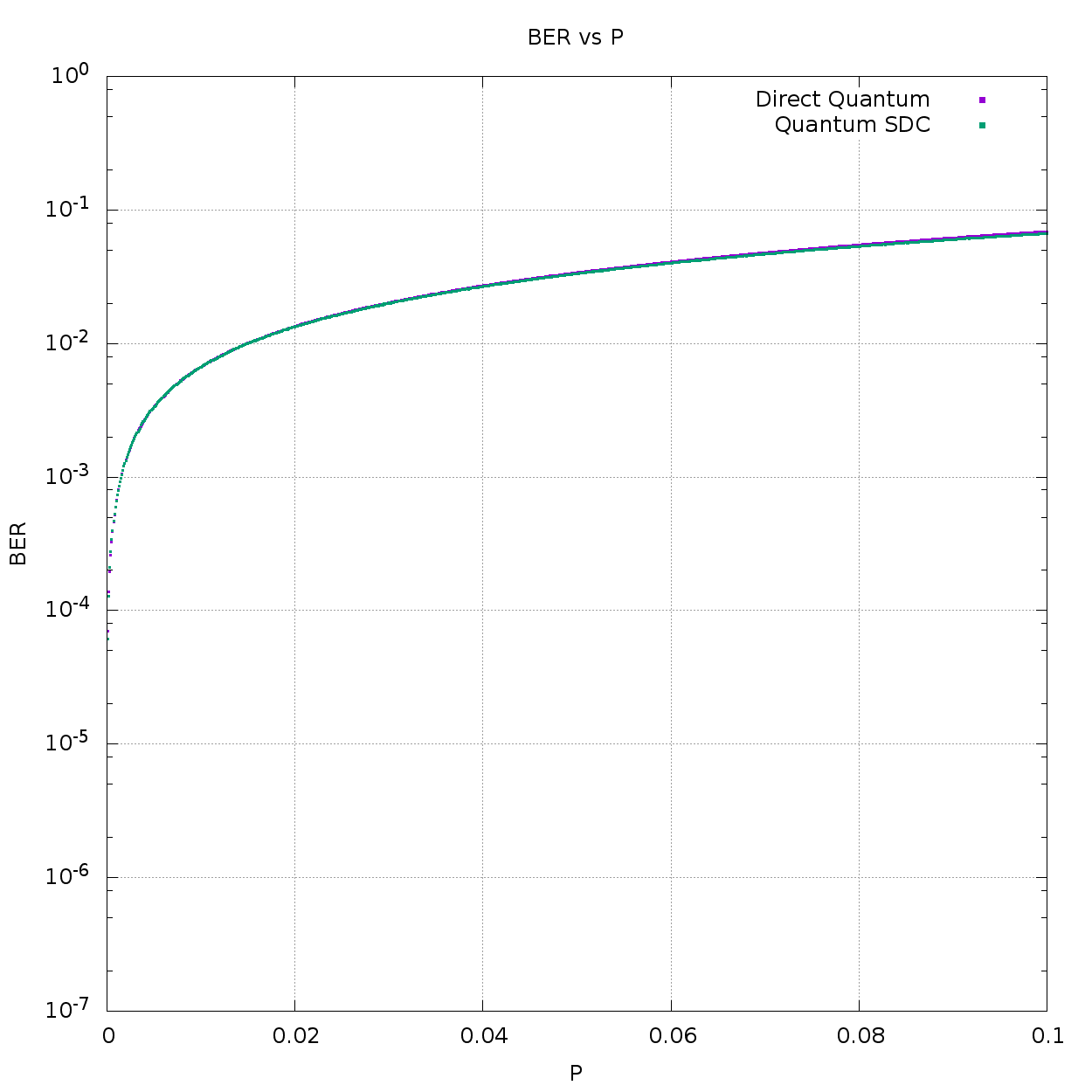}
\caption{BER average versus noise parameter $p$ for non-FEC encoded transmissions through a depolarizing channel. Points correspond to single-sided superdense coding (SDC) and direct transmissions of single qubits. }
\label{fig:noecc_ber}
\end{figure}

We calculate the BER after simulated transmission by performing a comparison between the decoded bit stream and the original input message. We use a nominal input of approximately 1 million bits, where slight adjustments are made to accommodate the number of bits required per classical FEC symbol. The baseline case of SDC transmission through the depolarizing noise channel with no FEC coding is shown in Fig.~\ref{fig:noecc_ber} alongside the case of direct quantum transmission. As expected from our earlier results, the rate at which errors occur is the the same for both protocols.
\par
\begin{figure}[ht]
\centering
\includegraphics[width=1.0\columnwidth]{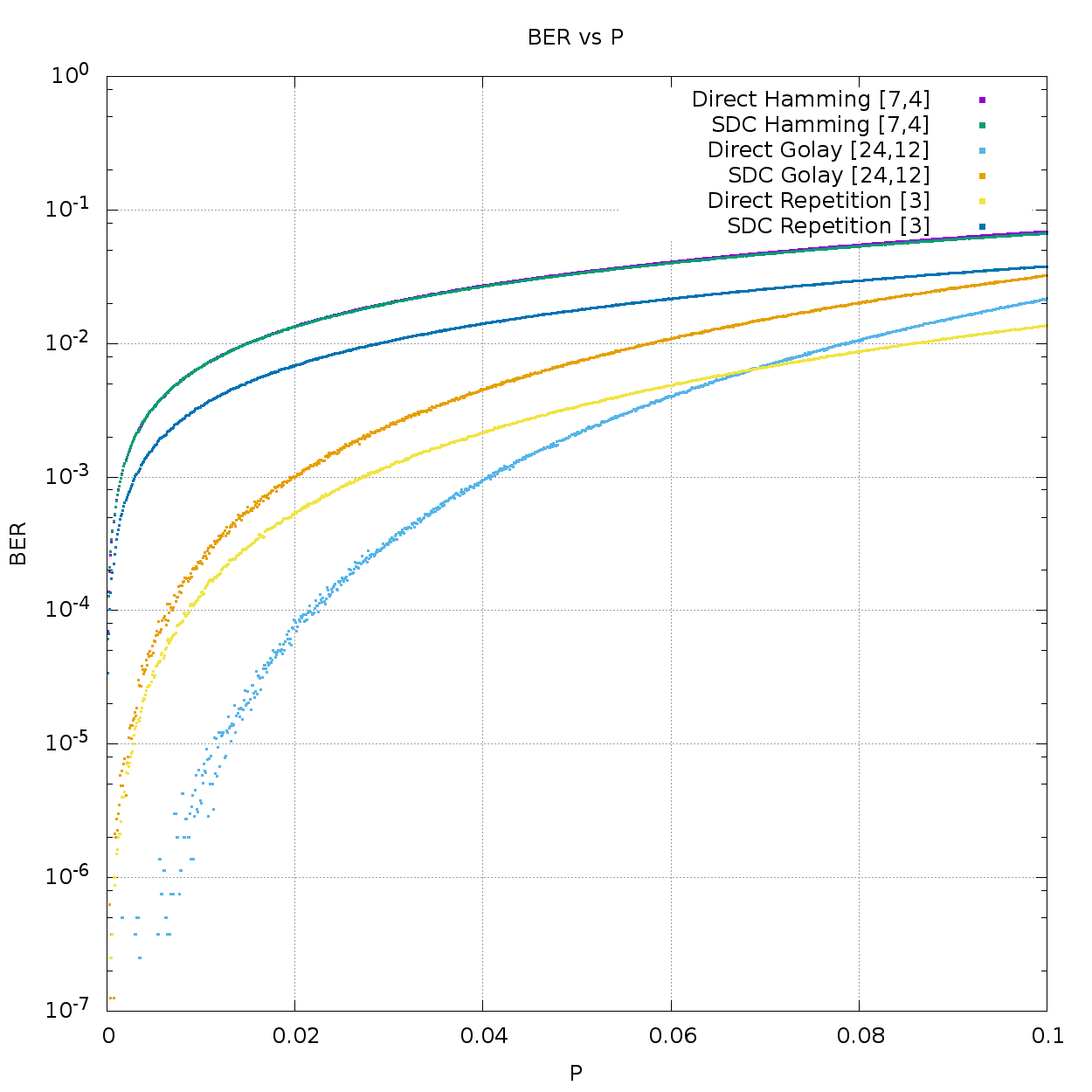}
\caption{BER averages versus noise parameter $p$ for FEC-encoded, non-interleaved transmission with either SDC or direct transmission. Curves correspond to Hamming [7,4], Golay [24,12], and repetition [3,1] codes.}
\label{fig:noninterleaved_ber}
\end{figure}

In comparison, simulation results for the average of FEC encoded transmissions are presented in Fig.~\ref{fig:noninterleaved_ber}. These curves show distinct behavior for the three types of FEC codes considered in Table~\ref{tab:codes}, which all perform better than the unencoded transmission schemes. The Golay code, which can correct up to weight-3 errors, performs the best when transmitting by SDC. However, there is a crossover with the direction transmission scheme using a repetition code. The Hamming code produces a similar curve for both SDC and direct transmission. Scatter in the plots for small values of $p$ are due to the increases in BER variance that arise from finite sampling.
\par
In Fig.~\ref{fig:noninterleaved_ber}, the behavior of the SDC and direct communication schemes are notably different even though they are both subject to the same noisy channel parameters. The reason for this difference traces back to the effect of the Pauli noise operators on the Bell states. In particular, the transmission of FEC codewords under depolarizing noise is susceptible to binary errors with weight greater than 1. For the single-sided Pauli noise mode, these errors arise when a transmitted Bell-state symbol is mapped into its exact opposite in terms of the 4-ary encoding. Moreover, these Pauli error operators yield correlated bit errors with respect to the subsequent decoding. However, many FEC codes work best with an independent random error model and the presence of correlated errors can undermine performance. This under performance results in a higher BER than arises for direct transmission, which cannot incur such correlated errors.
\par
\section{Interleaving}
\label{sec:interleaving}
In classical communication, data interleaving is frequently used to mitigate against correlated errors, also known as burst errors. Prior to transmission, binary elements of codewords output from the FEC encoder are interleaved with respect to transmission order while the received symbols are deinterleaved in the reverse order.  Although burst errors still occur during transmission, any correlated errors are effectively dispersed across the multiple, interleaved codewords. 
\par
As an example of interleaving, consider a two-bit input message $\{b_0, b_1\}$ protected by a repetition [3, 1] code. Encoding generates the binary string $\mathbf{c} = \{ b_0, b_0, b_0, b_1, b_1, b_1 \}$ which maps to the joint quantum state 
\begin{equation}
\rho_{\mathbf{c}} = \rho_{(b_0, b_0)} \kron \rho_{(b_0, b_1)} \kron \rho_{(b_1, b_1)}
\end{equation}
The simulation of transmission of this sequence using SDC through the depolarizing channel samples all 64 possible combinations of Pauli noise operators. After detection and decoding, the resulting BER is
\begin{equation}
\mathrm{BER}_{[3,1]} = \frac{1}{3}p + \frac{7}{18}p^2 + \frac{2}{27}p^3
\end{equation}
which has a leading order term that is linear in the noise parameter $p$.
\par
\begin{figure}[ht]
\centering
\includegraphics[width=1.0\columnwidth]{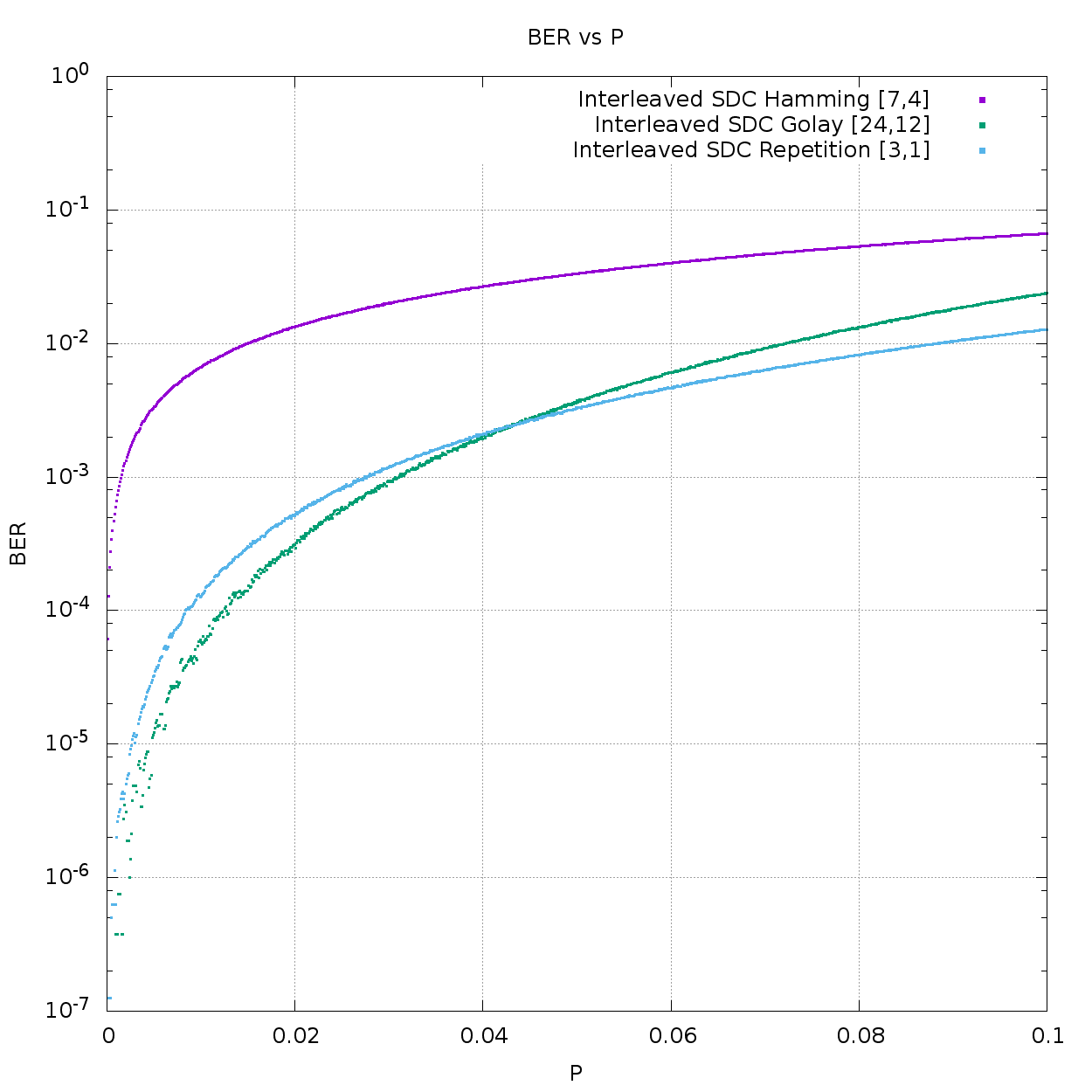}
\caption{BER average versus noise parameter $p$ for FEC-encoded, interleaved SDC transmission. Points correspond to FEC using Hamming [7,4], Golay [24,12], and repetition [3,1] codes.}
\label{fig:interleaved_ber}
\end{figure}

We contrast this non-interleaved result with a transmission scheme in which the output of the encoder is interleaved. In the current example, we will consider an interleaving pattern that alternatives between the first and second codewords, i.e., we prepare the interleaved binary sequence $\mathbf{c}' = \{ b_0, b_1, b_0, b_1, b_0, b_1 \}$,
which is mapped into the joint quantum state
\begin{equation}
\rho_{\mathbf{c'}} = \rho_{(b_0, b_1)} \kron \rho_{(b_0, b_1)} \kron \rho_{(b_0, b_1)}
\end{equation}
Although the noisy channel operator is the same, the probability for a burst error is greatly decreased by the interleaving and the overall BER is
\begin{equation}
\mathrm{BER}_{[3,1]}^{interleaved} = \frac{4}{3}p^2-\frac{16}{27}p^3
\end{equation}
In particular, the leading order term is now quadratic in the noise parameter $p$, which is an improvement over the non-interleaved results. Comparative plots of the BER for interleaved and non-interleaved cases are shown in Fig.~\ref{fig:interleaved_ber}.
\begin{figure}[ht]
\centering
\includegraphics[width=1.0\columnwidth]{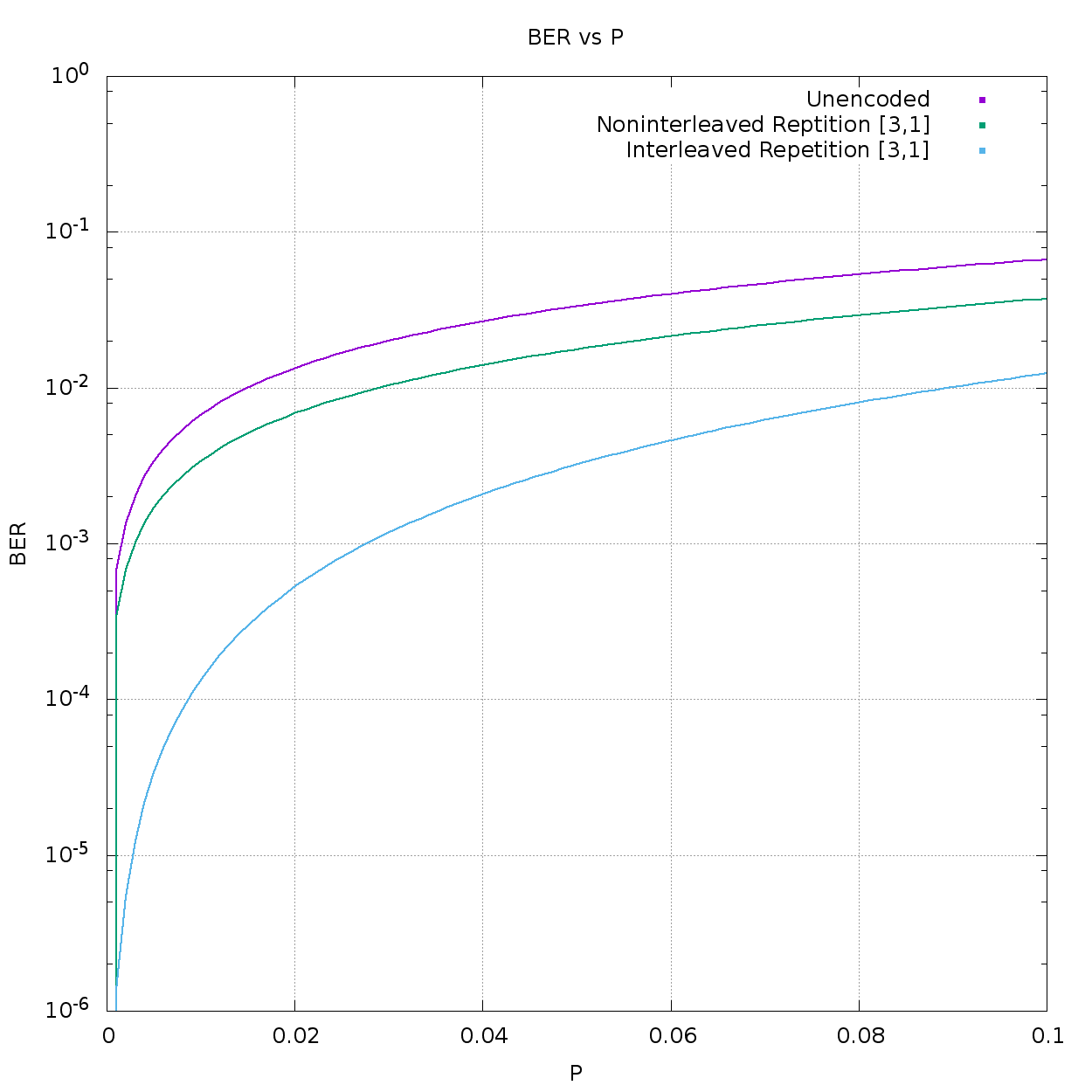}
\caption{BER average versus noise parameter $p$ for repetition encoded transmission with and without interleaving. The non-FEC encoded transmission is plotted for comparison.}
\label{fig:sdcquantum_repetition3_analytical_comparison}
\end{figure}
\par
We also present results for the BER obtained using interleaving for the Hamming and Golay codes in Fig.~(\ref{fig:interleaved_ber}).
\section{Conclusion}
\label{sec:conclusions}
We have presented a numerical study of how classical forward error correction techniques can be used with superdense coding. We have calculated the bit error rates for one-sided SDC using a Pauli channel model. Our simulations show the relative change in the BER when using Hamming, Golay, and repetition codes for both direct and superdense coding transmissions with respect to channel noise. These results are explained in terms of the varying distance of the FEC codes as well as the correlated noise that arises within the context of superdense coding. We have studied how to mitigate the influence of the correlated errors by using interleaving, in which reordering of the transmitted data is found to improve the ability of the code to protect against noise. This is shown to be due to the combined effects of Pauli errors on FEC codewords and the decoding method.
\par
Our results for the BER confirm the behavior expected for the entangled channel in the presence of isotropic Pauli noise, i.e., depolarizing noise. The effective bit-error rate is found to be $2p/3$ with $p$ the channel error rate. However, the effectiveness of the error correction depends subtly on the specific encoding and interleaving method used during transmission. For example, when $X$ errors on a transmitted entangled state lead to single-bit errors then $Z$ errors will yield weight-two errors on the classical message. The relative significance of these errors depends not only on the distance of the FEC code but also the position of the errors in the codeword. We have shown that interleaving is effective for mitigating against those higher-weight errors that would otherwise lead to uncorrectable states of the code.
\par 
Our use of classical FEC was motivated by a need to improve the reliability of quantum communication. While future QEC codes are likely to offer more error correction benefits, they will also require higher dimensional entangled quantum resources. Current limitations on quantum state preparation represent a bottleneck for using QEC to improve the quality of service in quantum communication systems. We have shown that FEC codes with proper interleaving offers an effective method that may be immediately employed with existing quantum communication hardware. We expect follow-on studies of other FEC codes will prove especially interesting for quantum communication systems, including polar codes for efficiency and erasure codes for additional noise models. 
\par
More generally, future applications for FEC in quantum communication include superdense coding as well as quantum key distribution, especially in the context of multi-user communication networks that may require codes specialized to different channel modes \cite{Grice2011}. The FEC methods may also be useful for improving the detection of unmodulated states in tamper-indicating quantum seals \cite{Williams2016}. However, we must expect that QEC codes will ultimately be necessary for any full-scale quantum networking environment. Protocols that do not communicate classical information should perform better with QEC than with FEC, since only QEC can be used to correct errors in the quantum state itself. This includes applications in multi-user quantum communication protocols that teleport quantum information \cite{Humble2010,Humble2011,Humble2011b} or swap entanglement between communication channels \cite{Humble2008}. A promising approach to reconfiguring networks between FEC and QEC encoding methods is to exploit programmable nodes \cite{Dasari2015}, in which the error correction scheme is determined by an external application based on channel diagnostics.
\par
This work was supported by the Defense Threat Reduction Agency Basic Research, the Army Research Laboratory, and the Department of Energy HERE program. This manuscript has been authored by UT-Battelle, LLC, under Contract No. DE-AC0500OR22725 with the U.S. Department of Energy. The United States Government retains and the publisher, by accepting the article for publication, acknowledges that the United States Government retains a non-exclusive, paid-up, irrevocable, world-wide license to publish or reproduce the published form of this manuscript, or allow others to do so, for the United States Government purposes. The Department of Energy will provide public access to these results of federally sponsored research in accordance with the DOE Public Access Plan.
\bibliography{paper}{}
\bibliographystyle{apsrev4-1}

\end{document}